\documentclass[12pt]{iopart}

\expandafter\let\csname equation*\endcsname=\relax
\expandafter\let\csname endequation*\endcsname=\relax
\usepackage{amsmath,amssymb,amsthm} %

\usepackage{graphicx}
\usepackage{xcolor}
\usepackage{mathtools}
\usepackage{hyperref}

\def\ket#1{\left|#1\right\rangle}

\def\scal#1#2{\langle#1|#2\rangle}

\def\abs#1{\left\lvert#1\right\rvert}

\def\abs#1{\left|#1\right|}

\usepackage{tikz}

\usepackage{makecell}

\newcommand{\der}{\operatorname{d\!}{}}

\newcommand{\ave}[1]{\left< #1 \right>}

\newcommand{\ii}{\mathrm{i}}

\usepackage{booktabs}
\newcommand{\ra}[1]{\renewcommand{\arraystretch}{#1}}

\newcolumntype{R}{>{$}r<{$}}
\newcolumntype{L}{>{$}l<{$}}
\newcolumntype{M}{R@{${}={}$}L}

\begin{document}

\title[A differentiable programming method for quantum control]{A differentiable programming method\\ for quantum control}

\author{Frank Sch\"{a}fer, Michal Kloc, Christoph Bruder, Niels  L\"{o}rch}

\address{Department of Physics, University of Basel, Klingelbergstrasse 82, CH-4056 Basel, Switzerland}
\ead{frank.schaefer@unibas.ch, michal.kloc@unibas.ch}
\vspace{10pt}
\begin{indented}
\item[]\today
\end{indented}

\begin{abstract}
Optimal control is highly desirable in many current quantum systems, especially to realize tasks in quantum information processing.
We introduce a method based on differentiable programming to leverage explicit knowledge of the differential equations governing the dynamics of the system. 
In particular, a control agent is represented as a neural network that maps the state of the system at a given time to a control pulse. 
The parameters of this agent are optimized via gradient information obtained by direct differentiation through both the neural network \emph{and} the differential equation of the system.
This fully differentiable reinforcement learning approach ultimately yields time-dependent  control parameters optimizing a desired figure of merit. 
We demonstrate the method's viability and robustness to noise  in eigenstate preparation tasks for three  systems: a~single qubit, a~chain of qubits, and a quantum parametric oscillator. 
\end{abstract}

%
\vspace{2pc}

\section{Introduction}
In many applications, effective manipulation of physical systems requires an optimization of the available control parameters such as control fields.
Since classical intuition often fails for quantum mechanical systems, devising optimal control strategies can be very challenging~\cite{dong2010}.
A wide range of traditional  optimization methods obtain the optimal control pulse sequences by a gradient-based maximization of a certain, task-specific objective~\cite{glaser2015, krotov89, khaneja05,chen14,chen15, morzhin18}.
Additional constraints beside the main objective can be efficiently  implemented using automatic differentiation (AD)~\cite{leung17}
and take advantage of GPU acceleration.

In recent years, astonishing advances in reinforcement learning (RL) has initiated great opportunities for control optimization.
In black-box RL, the training is based purely on collecting rewards and subsequent modification of network parameters to maximize the expected reward over all possible trajectories without any explicit representation of the system ~\cite{sutton1998}.
In contrast to standard optimal control algorithms, perturbations of the initial state are naturally captured within this framework.

Successful demonstrations of model-free RL in physics include, e.g., the manipulation of a quantum-spin chain~\cite{bukov18PRX}, the inversion of the quantum Kapitza oscillator~\cite{bukov18PRB}, and  quantum gate-control optimization~\cite{niu2019}.
The authors of Ref.~\cite{bukov18PRX} also show that the performance is comparable to standard optimal control algorithms.
By implementation of physical knowledge through a sophisticated reward scheme a student/teacher network could discover quantum error correction strategies from scratch~\cite{fosel2018}.

On the contrary, model-based RL approaches learn an explicit representation of the system simultaneously with a value function and policy.
The model of the environment is built from the (potentially sparse) reward signals which is, however, computationally expensive.
Nevertheless, if feasible, model-based RL bears the potential to learn optimal policies from a smaller number of environment interactions and efficiently handle changing objectives as well~\cite{atkeson1997, kurutach2018}.

Deep-learning experience shows that gradient-based approaches can be trained much more efficiently as the gradients indicate a useful search direction for optimization.
Despite the fact that real-world data cannot be differentiated, the associated simulations can be.
Recently, a great speed-up in the optimization process was demonstrated if the environment is implemented as a differentiable engine~\cite{degrave2019, Peres2018}.
Once the environment is differentiable, AD allows us to backpropagate directly through the time trajectory and we thus effectively switch from model-free to model-based RL~\cite{innes19b, duvenaud2018}.
This is a part of a new paradigm named differentiable programming (DP) where programs are designed in a fully differentiable way~\cite{innes19,rackauckas2020}.
The learnable structures, such as neural networks (NNs), are then embedded in a standard procedural code as a way to obtain the resulting program. 
Among the first examples, it was demonstrated that DP allows  to find optimal solutions much faster than model-free RL for standard classical problems~\cite{innes19b,innes19}.
In physics, tensor network algorithms were programmed in a fully differentiable way~\cite{liao19}.

In this paper, we follow this pivotal shift in the usage of RL and show that  using the DP method, a predictive model can be trained  to provide successful control strategies for noisy inputs within different quantum mechanical systems.
We start by introducing and illustrating the workflow of our method using the most basic quantum system of a single qubit in Section~\ref{Sec:Qubit}. We then evaluate the method on two further examples:
In Section~\ref{SubsecGHZ} we investigate the preparation of the GHZ state in a chain of qubits.
In Section~\ref{Subsec:EigPrepPar}, the eigenstate preparation in the case of a quantum parametric oscillator is considered.

\section{Quantum control problem}
\label{Sec:QControl}
Suppose a quantum system is described by a Hamiltonian 
\begin{equation}
  H(t)=H_0+\sum_{k=1}^K u_k(t)H_k\,,
  \label{Eq:ContHam}
 \end{equation}
where $H_0$ is a time-independent part (which we will refer to as the drift term) and $H_k$ represents control fields with respective scalar amplitudes $u_k(t)$.
The total number of independent control fields is $K$.
The dynamics of a state $\ket{\psi(t)}$, represented as a vector in a Hilbert space, is governed by the Schr\"{o}dinger equation $\ii \partial_t \ket{\psi(t)}= H(t) \ket{\psi(t)}$ given some initial state $\ket{\psi(t_0)}\equiv \ket{\psi(t=0)}$ (we use $\hbar=1$ throughout the paper).
The goal of quantum control is to optimize $u_k(t)$ over a given time interval according to a certain figure of merit, e.g., to maximize an overlap  with some target state.
The overlap $ F(t)=\abs{\scal{\psi(t)}{\psi_{\rm tar}}}^2$ is generally called fidelity.

Many numerical approaches consist of discretizing the time interval into $N$ steps in which the control amplitudes are considered constant.
The pulse sequences $u_k(t_i)$, $i=1\ldots N$ are then optimized by gradient methods.
Popular examples are the GRAPE algorithm~\cite{khaneja05} and the class of algorithms generally attributed to Krotov~\cite{krotov89}.
Alternative approaches like the CRAB algorithm~\cite{Doria11,caneva2011} do not rely on evaluating gradients but rather map the problem to finding the extrema of a multivariate function using direct search methods  such as Simplex methods~\cite{Doria11}.

The standard control schemes are not generally equipped to deal with uncertainties in the initial state.
If the input state is altered, the control fields will in general no longer be optimal.
It should be noted, though, that proposals for generalization of the approach have been already put forward, see Refs.~\cite{chen14,wu18,wu19}. 
RL formulates control tasks in terms of Markov Decision Processes which map an individual state $\ket{\psi(t_i)}$ to the associated actions. 
Therefore, the noisy inputs can be treated very naturally, provided that the agent was exposed to sufficiently many training examples.
Our differentiable programming method, as described in the next section, leverages a similar strategy.

\section{Quantum optimal control with differentiable programming}
\begin{figure}[h!]
	\centering
	\includegraphics[width=1\linewidth, angle=0]{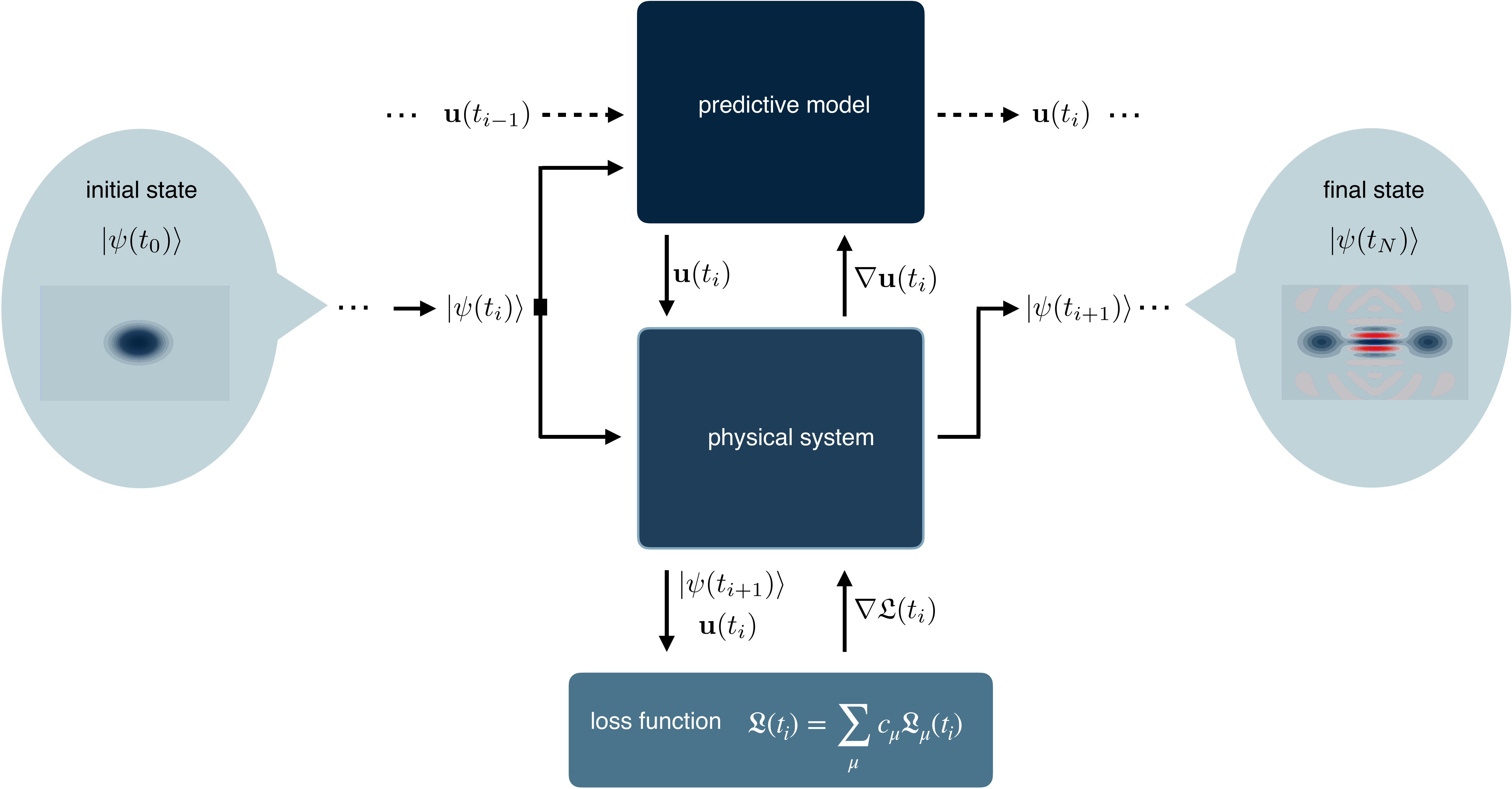}
	\caption{Sketch of our differential programming approach for quantum optimal control. 
	In the forward pass, starting from an initial state, $\ket{\psi(t_0)}$ (and vanishing control fields), in each time step $i$, the predictive model maps the quantum state $\ket{\psi(t_i)}$ to the associated optimal control amplitudes 
    ${\bf u}(t_{i})$. 
    These amplitudes determine the time evolution of the physical system from $\ket{\psi(t_i)}$ to $\ket{\psi(t_{i+1})}$, as computed by an ODE solver. 
    Both the state $\ket{\psi(t_{i+1})}$ and the amplitudes ${\bf u}(t_{i})$ contribute to the loss function  $\mathfrak{L}(t_i)$.
    This procedure is iterated for all $N$ steps. 
    In the backward pass, automatic differentiation allows us to compute the gradients of $\mathfrak{L}(t_i)$ with respect to the parameters of the predictive model by direct differentiation through the ODE solver. 
    These gradients are used to optimize the model's parameters in a series of epochs. To ensure sufficiently smooth control fields and to achieve a faster learning progress, we also give the last amplitudes ${\bf u}(t_{i-1})$ to the predictive model as an input (dashed arrows). }
	\label{fig_difprog}
\end{figure}

Let us now describe how a differentiable programming approach allows us to solve the quantum optimal control problem as defined in Section~\ref{Sec:QControl}. 
The method consists of three main building blocks: the predictive model, the physical system, and the loss function~$\mathfrak{L}$.
The connection between these building blocks is depicted in Fig.~\ref{fig_difprog}. 
Starting from some initial state, $\ket{\psi(t_0)}$, and vanishing control fields, in each time step $i$ the predictive model receives the current quantum state, $\ket{\psi(t_i)}$, as well as the last control field strengths, $u_k(t_{i-1})$, as an input. 
The predictive model maps this input to the next control field strengths, $u_k(t_{i})$. 
Adopting the terminology of RL, we now also refer to the predictive model as \textit{agent} and to the control fields as \textit{actions}. 
According to our experience,  if the agent is aware of its last actions, the optimal strategy can be found more easily and often provides smoother control signals. Moreover, the output of the predictive model can be easily reformulated to return  the  gradients rather than the action themselves. In that case, the maximal fluctuations of the control fields between the time steps can be naturally controlled which may be required in experimental applications.
Solving the ordinary differential equation (ODE) of the physical system, as determined by the Hamiltonian and the Schr\"odinger equation, leads to the next state  $\ket{\psi(t_{i+1})}$. 
This state, as well as the  actions taken, $u_k(t_{i})$, enter the computation of the loss function which maps the inputs to a scalar value. 
This procedure is iterated for all time steps. 

\begin{figure}[h!]
	\centering
	\includegraphics[width=1\linewidth, angle=0]{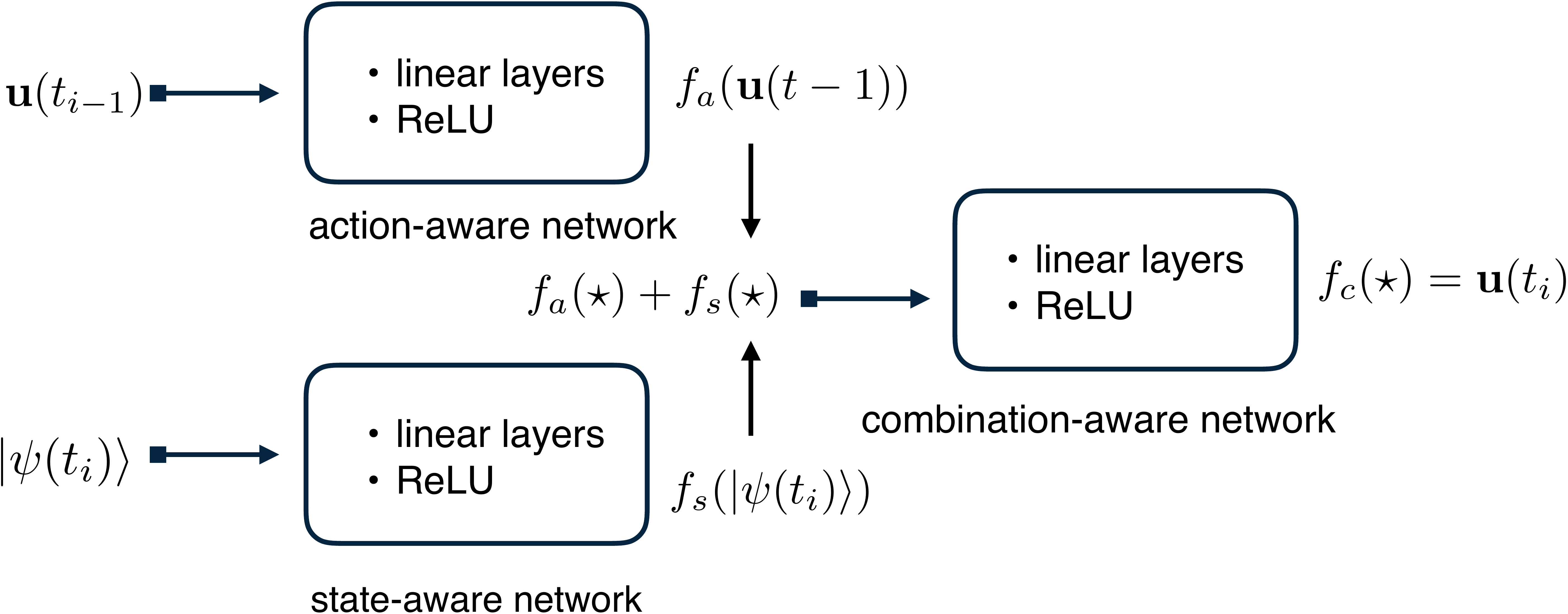}
	\caption{The general neural network architecture for the predictive model used throughout all our numerical experiments. Quantities in square brackets indicate the respective input/output tensor dimensions. The number of weights and layers used for each network is given in Tab.~\ref{TabHyperPars} in~\ref{Ap:Table}. }
	\label{fig_arch}
\end{figure}

Figure~\ref{fig_arch} shows the architecture of the predictive model.
We use NNs with linear layers and rectified linear units (ReLUs) as the activation functions. 
We divide the full NN  into three separate networks:
\begin{enumerate}
	\item A \textit{state-aware} network $f_s$ that takes the current quantum state, $\ket{\psi(t_i)} \in \mathbb{C}^{D}$, and computes a map 
	\begin{equation}
	f_s(\ket{\psi(t_i)}): \mathbb{C}^{D} \rightarrow  \mathbb{R}^{B},
	\end{equation}  
	where $D$ is the Hilbert space dimension and $B$ is the number of output features in the final layer of the state-aware network. 
	
	\item An \textit{action-aware} network $f_a$ that takes the last actions, $u_k(t_{i-1}) \in \mathbb{R}^{K}$, and computes a map 
	\begin{equation}
	f_a(u_k(t_{i-1})): \mathbb{R}^{K} \rightarrow  \mathbb{R}^{B},
	\end{equation}  
	where $K$ is the total number of control fields and $B$ is identical as above, and
	
	\item a \textit{combination-aware} network $f_c$ that post-processes the sum of the two maps $f_s(\ket{\psi(t_i)})+f_a(u_k(t_{i-1}))$, and predicts the next control fields
	\begin{equation}
	f_c(f_s(\ket{\psi(t_i)})+f_a(u_k(t_{i-1}))): \mathbb{R}^{B} \rightarrow  \mathbb{R}^{K},
	\end{equation}  
	with $B$ and $K$ as above.
	
\end{enumerate}
Here, we combined $f_s$ and $f_a$ by a simple addition, but other combinations such as concatenation are also possible.  Which of the combinations is most efficient  in  a  given  situation  will  depend  on  the  particular physical system and number of parameters of the model.

All our code is implemented in PyTorch~\cite{paszke2017} which has no complex tensor support, yet. Therefore, we map the Schr\"odinger equation via the isomorphism 
\begin{equation}
\partial_t \ket{\psi(t)} = -\ii H(t) \ket{\psi(t)} \xmapsto {} \partial_t \begin{bmatrix}
  \ket{\psi_{\rm Re}(t)}\\
 \ket{\psi_{\rm Im}(t)}	
\end{bmatrix}=   \begin{bmatrix}
H_{\rm Im}(t)& H_{\rm Re}(t)\\
- H_{\rm Re}(t)& H_{\rm Im}(t)
\end{bmatrix}
\begin{bmatrix}
\ket{\psi_{\rm Re}(t)}\\
\ket{\psi_{\rm Im}(t)}	
\end{bmatrix}
\label{eq:ODE-system}
\end{equation}
to real-valued vectors and matrices.
The subscripts $\rm Re/Im$ denote the real and imaginary part, respectively.
Similarly, $\ket{\psi(t_i)}$ is mapped onto $\left[	\ket{\psi_{\rm Re}(t_i)}, \ket{\psi_{\rm Im}(t_i)}\right]^T$ as the input for the state-aware network.	
For the time evolution of the physical system, as described by the ODE problem \eqref{eq:ODE-system}, we use Heun's method ~\cite{suli2003} to get the next time point $i+1$.

The subsequent computation of the loss function consists of a weighted sum of individual contributions $\mathfrak{L}_\mu$  that have a well-defined physical meaning. These terms are explained in detail in Section~\ref{Sec:Loss}. 
Importantly, since the physical system is implemented entirely in PyTorch and  is differentiable, the derivative of the loss function  with respect to the network's weights can be computed at any point in time.

We use the Adam optimizer~\cite{kingma2014} to train the model in a series of epochs, where $b$ simulations are carried out in parallel per epoch, see Fig.~\ref{fig_arch}. 
Hyperparameters, such as the coefficients of the individual loss terms as well as the number of  output features of the linear layers, are obtained either empirically or using the Bayesian Optimization and Hyperband method (BOHB)~\cite{falkner2018} which is a combination of random search (bandit-based method) with Bayesian optimization.

The embedding of the physical model into the backpropagation pass (which is not the case in model-free RL), has led to far more effective control strategies and faster training for classical environments~\cite{innes19b}.
The potential robustness of the method with respect to noise in the input is desired for practical purposes where inputs can not be prepared to arbitrary precision.  
To test the performance of our scheme, we apply it to different physical systems, which are introduced in the next section.

\section{Physical systems}
\subsection{Qubit chain with nearest-neighbor interaction}
\label{Subsec:TheoryQubit}
\begin{figure}
	\centering
  \includegraphics[width=1\linewidth, angle=0]{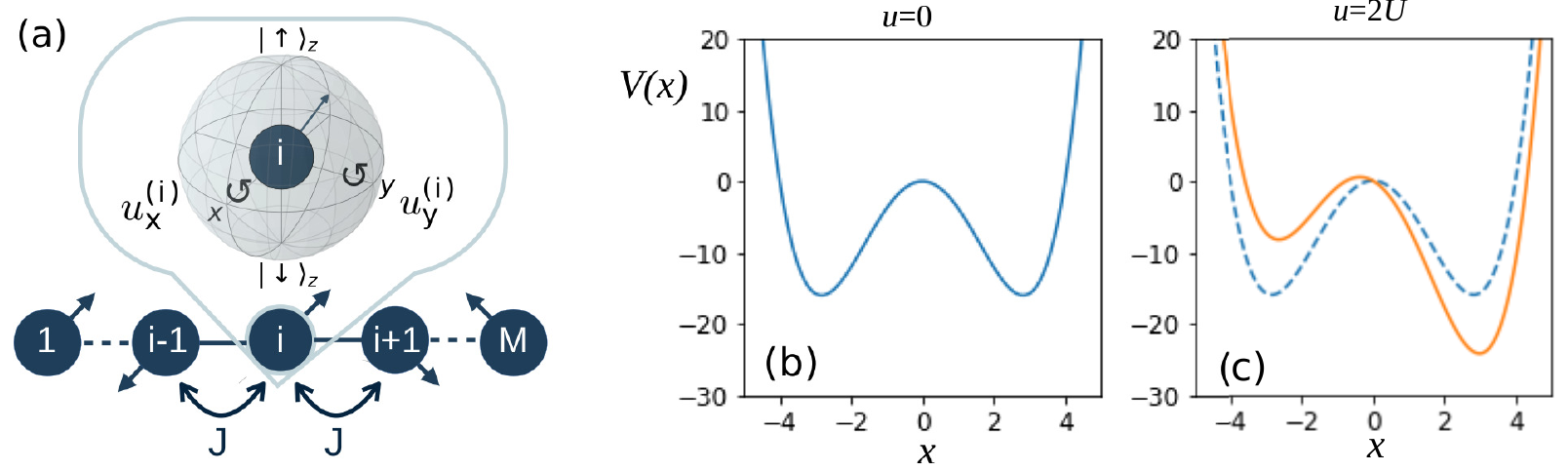}
	\caption{Panel (a): A chain of $M$ qubits and the control fields. According to the Hamiltonian~\eqref{Eq:SpinHam}, each qubit interacts with its  nearest neighbors with strength $J$.
	The control fields $u_x^{(i)}, \ u_y^{(i)}$ are applied at each site.
	Therefore, any qubit can be rotated by an angle prescribed by the respective field strengths.
	Panels (b)~and~(c): Shape of the classical potential of the Hamiltonian~\eqref{Eq:Parametron} describing a quantum parametric oscillator. 
	Both cases with the control field $u$ being switched off (panel b) and switched on (panel c) are depicted. 
	}
	\label{fig_spinchain}
\end{figure}

We consider a chain of $M$ identical qubits (or spins $\frac{1}{2}$) with the control Hamiltonian 
\begin{equation}
    H(t) = \sum_{i=1}^{M} J  \sigma^{(i)}_z\sigma^{(i+1)}_z + u_x^{(i)}(t) \sigma_x^{(i)}+u_y^{(i)}(t) \sigma_y^{(i)}\,,
    \label{Eq:SpinHam}
\end{equation}
where $\sigma^{(i)}_x, \sigma^{(i)}_y,\sigma^{(i)}_z$ are the Pauli matrices acting on the $i$th site of the chain.
For any single qubit we have two independent control fields with amplitudes $u_x^{(i)}(t)$ and $u_y^{(i)}(t)$.
They allow us to rotate each spin by an arbitrary angle as illustrated in Fig.~\ref{fig_spinchain}(a).

The dimension of the Hilbert space scales as $D=2^M$.
The set of vectors representing all possible configurations of the $z$ projection at individual sites forms a basis.
Therefore a general state vector $\ket{\psi}$ can be expanded as
\begin{equation}
\ket{\psi}=C_0 \ket{\uparrow \uparrow \uparrow .. \uparrow \uparrow }_z+C_1 \ket{\uparrow \uparrow \uparrow ..\uparrow \downarrow }_z+\ldots+C_{D-1} \ket{\downarrow \downarrow \downarrow .. \downarrow \downarrow }_z\,,
\label{Eq:GenVecSpinChain}
\end{equation}
with unique coefficients $C_0,\ldots,C_{D-1}$.
The drift term has the form of a nearest-neighbor interaction.
As we set $J>0$, the ground state is given by the doubly degenerate manifold spanned by the N\'{e}el states $\ket{\psi_{\rm gs}^1} = \ket{\uparrow \downarrow \uparrow \ldots}_z $ and $ \ket{\psi_{\rm gs}^2} = \ket{ \downarrow \uparrow \downarrow \ldots}_z $.

In Section~\ref{Sec:Qubit} we consider a simplified model case with only one qubit and the Hamiltonian
\begin{align}
H = \frac{\omega}{2} \sigma_z + u_x(t) \sigma_x\,,
\label{eq_qubit}
\end{align}
where $\omega$ is the qubit frequency.
Note that the single control field $u_x(t)$ is sufficient to realize any rotation as $\sigma_y$ can be obtained as a commutator of $\sigma_x$ and $\sigma_z$.

\subsection{Quantum parametric oscillator}
 The other system considered is a driven non-linear optical resonator whose dynamics can be described by the following Hamiltonian in the rotating frame (see, for example, Ref.~\cite{bartolo2016})
\begin{equation}
H(t)=U a^\dag a^\dag a a +G\left(a^\dag a^\dag +a a\right) +u(t)\left(a+a^\dag \right)\,.
\label{Eq:Parametron}
\end{equation}
The first term is a Kerr (non-linear) interaction between the photons.
They are annihilated and created using the respective operators $a, \ a^\dag$.
We can represent a general state $\ket{\psi}$ in the Fock basis, i.e., the basis of photon occupation number states $\ket{n}$, as
\begin{equation}
\ket{\psi}=\sum_{n=0}^{\infty}C_n\ket{n}, \qquad C_n\equiv\scal{n}{\psi}\,.
\label{Eq:ParaState}
\end{equation}
The mean photon number in the state $\ket{\psi}$ is given as $\ave{n}_\psi = \scal{\psi}{a^\dag a|\psi}$.

In the following, we fix the amplitude of the coherent two-photon drive relative to the strength of the Kerr non-linearity $G=-4U$.
Our control scheme consists of one tunable parameter $u$  which controls the amplitude of the single-photon drive.
Its action can be intuitively understood from the classical counterpart of the system.
The classical potential $V(x)$ is obtained from Eq.~\eqref{Eq:Parametron} by writing the operators as $c$-numbers $\sqrt{2}a=x+\ii p$ and setting $p=0$.
Figure~\ref{fig_spinchain}(b) reveals a symmetric double-well potential when the control field is switched off.
If the control field is switched on, the potential becomes tilted as shown in Fig.~\ref{fig_spinchain}(c).

\section{Loss functions}
\label{Sec:Loss}
Multiple objectives can be passed on to the agent via  a  case-specific loss function $\mathfrak{L}$.
We assume the generic form~\cite{leung17,caneva2011,abdelhafez19}
  \begin{equation}
  \mathfrak{L}=\sum_{\mu} c_\mu \mathfrak{L}_\mu \,,
  \label{Eq:FundLoss}
  \end{equation} 
which is a linear combination of elementary loss functions $\mathfrak{L}_\mu$ encoding specific details of the control task. 
The agent will reflect these details according to their relative importance given by the coefficients $c_\mu$. 
They have to be optimized either empirically (as in Ref.~\cite{leung17}) or by means of some hyperparameter optimization algorithms.

\begin{table*}[h!]
	\centering
	\begin{tabular}{llM}\toprule
	\multicolumn{4}{l}{primary goal }  \\
	& target-state infidelity & \mathfrak{L}_{F}	&  \sum_{i}^{N}{\gamma^i \left(1-\abs{\scal{\psi(t_i)}{\psi_{\rm tar}}}^2\right)}  	 \\
	& & \mathfrak{L}_{FN} & 1- \abs{\scal{\psi(t_N)}{\psi_{\rm tar}}}^2   \\
	\multicolumn{3}{l}{constraints}  \\	
	& control amplitudes & \mathfrak{L}_{\rm amp} & \sum_{i}^{N}{\abs{{\bf u} (t_i)}} \\
	& & \mathfrak{L}'_{\rm amp} & \sum_{i}^{N}{\abs{ {\bf u}(t_i)}^2} \\
		\bottomrule
	\end{tabular}
	\caption{Elementary loss functions employed for quantum optimal control in our numerical experiments.}
	\label{TabFundLoss}
\end{table*}

The elementary loss functions used in this paper are summarized in Tab.~\ref{TabFundLoss}.
The function $\mathfrak{L}_{F}$ is a  discounted sum over the infidelity for every time step $i$ with the real positive discount factor $\gamma \le 1$. 
This loss function serves as the main objective for our control problems as its minimization leads to a maximal overlap with the target state on the entire control interval. Therefore, minimization of $\mathfrak{L}_{F}$ minimizes the time to reach the target state.

In the case of eigenstate preparation, once the target state has been reached, the control fields can be switched off.
Then, the state remains unchanged (up to a complex phase) under the time evolution according to the drift Hamiltonian.
We add the loss function $\mathfrak{L}_{FN}$ to ensure that the target state is prepared at the final time step.
To focus on another particular point in time, $\mathfrak{L}_{FN}$  can be adjusted accordingly.

The remaining loss functions in Tab.~\ref{TabFundLoss} represent additional constraints on the final control pulse sequence.
Namely, by employing $\mathfrak{L}_{\rm amp}$ or $\mathfrak{L}'_{\rm amp}$ the agent is forced to prefer smaller amplitudes.\footnote{Here, the difference is purely technical, however, the squared amplitude can be  linked with the intensity of the control field which in many cases represents an experimentally more relevant quantity than the amplitude itself.} 
Of course, other elementary loss functions can be added to Eq.~\eqref{Eq:FundLoss} according to the specific requirements of the control task.

\section{Numerical experiments}
\subsection{Optimal control of a single qubit}
\label{Sec:Qubit}

We start with a minimal quantum  model to illustrate the workflow of our method.
We consider a qubit in a magnetic field that can be manipulated by a control field in $x$ direction described by the Hamiltonian~\eqref{eq_qubit}.
The general task is to move an arbitrary initial state $\ket{\psi(t_0)}= \cos(\frac{\theta}{2}) \ket{\uparrow}_z+ \sin( \frac{\theta}{2}) e^{i\phi}\ket{\downarrow}_z$, with $\theta \in [0, \pi]$ and $\phi \in [0,2\pi)$ uniformly sampled on the Bloch sphere, into the state $\ket{\psi_{\rm tar}} =\ket{\uparrow}_z$ with high fidelity in the shortest possible time. 
Note that this task is significantly different from the typical quantum optimal control setup where the initial state has fixed  values for $\theta$ and $\phi$. 

The predictive model is constructed as in Fig.~\ref{fig_arch} and the number of weights and biases are constant for all qubit experiments, see Tab.~\ref{TabHyperPars} in~\ref{Ap:Table}.
We use the Bayesian Optimization and Hyperband (BOHB) method~\cite{falkner2018} to tune the number of parallel simulations $b$, the learning rate of the Adam optimizer, and the coefficients $c_F, \ c_{FN}$ and $c'_{\rm amp}$ of the loss function  $\mathfrak{L}$~\eqref{Eq:FundLoss}. 
The objective for BOHB is not allowed depend explicitly on the $c_\mu$, as they are also optimized.
A natural choice for the BOHB objective is $\mathfrak{L}_F$ from Tab.~\ref{TabFundLoss}.
More details  of the hyperparameter optimization can be found in~\ref{Ap:BOHB}.

\begin{figure}
	\centering
	\includegraphics[width=1\linewidth, angle=0]{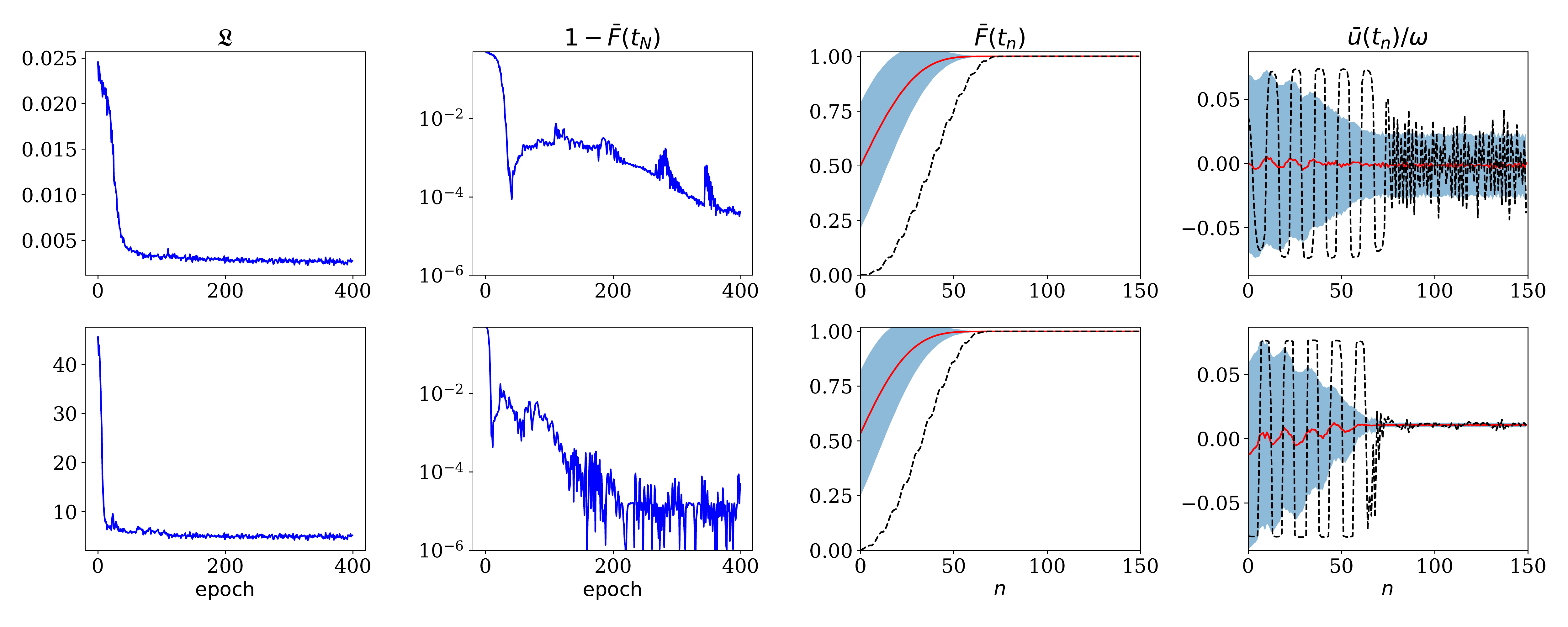}
	\caption{ 
		Preparation of the eigenstate  $\ket{\uparrow}$ of the drift term of the Hamiltonian~\eqref{eq_qubit}. 
		The first row shows the results when the loss function $\mathfrak{L}$~\eqref{Eq:FundLoss} contains only one non-zero
		coefficient $c_F$. 
		The second row shows the optimized results for three contributing coefficients $c_F$, $c_{FN}$,
		$c_{\rm amp}$. 
		In the first column, the evolution of $\mathfrak{L}$ during the training phase is shown. The second column shows the mean value of the final state infidelity as a function of the epochs. 
		The third and fourth columns show the performance of the model when it is applied to a test set of
		size 512. 
		The red curve/ blue shaded region indicate the mean/standard deviation of the fidelity and the actions, respectively.  
		The black dashed line highlights the special case with initial state given as $\ket{\psi(t_0)}=\ket{\downarrow}$. 
		 All hyperparameters can be found in Tab.~\ref{TabHyperPars} in~\ref{Ap:Table}.}
	\label{fig_qubit}
\end{figure}

The results of the control task with the optimized hyperparameters are shown in Fig.~\ref{fig_qubit}. 
The two rows compare different choices of the number of hyperparameters in the loss function~\eqref{Eq:FundLoss}.
The first two columns visualize the loss and the mean of the final state infidelity as a function of the epochs. 
Smaller final fidelities are reached when several loss functions are combined.
We applied the trained model to a test set of 512 random configurations.
The results for the mean and the standard deviation of the fidelity  as a function of time and of the applied control fields clearly show the different behaviors for different choices of hyperparameters in the loss function.
Particularly, the term $\mathfrak{L}'_{\rm amp}$ pushes the actions to zero and hence helps to achieve a well-defined shorter control sequence.
Comparison with previous results  \cite{leung17} shows that we get a very similar control pulse when the initial state is chosen as the ground state of the drift term of the Hamiltonian~\eqref{eq_qubit}.
A comparison with a RL approach can be found in~\ref{Ap:CompareRLvsDP}.

\subsection{GHZ state preparation in a chain of qubits}
\label{SubsecGHZ}
In this section, we aim at the preparation of a GHZ state~\cite{greenberger1989} in a chain of $M$ qubits
\begin{equation}
\ket{{\rm GHZ}}= \frac{1}{\sqrt{2}} \left( \ket{\uparrow}_z^{\otimes M} + \ket{\downarrow}_z^{\otimes M} \right) =  \frac{1}{\sqrt{2}} \left( \ket{\uparrow \uparrow \dots}_z + \ket{\downarrow \downarrow \dots}_z \right)\,.
\end{equation} 
These states play an important role, e.g., for multi-particle generalizations  of  superdense  coding~\cite{bose1998} or in quantum secret sharing~\cite{hillery1999}.

The initial state $\ket{\psi(t_0)}$ is chosen as one of the ground states of the drift Hamiltonian from Eq.~\eqref{Eq:SpinHam} (states $\ket{\psi^1_{\rm gs}}$ and $\ket{\psi^2_{\rm gs}}$, see Section~\ref{Subsec:TheoryQubit}).
We assume the input contains noise which is implemented as flipping any single qubit in the initial state with $10\%$ probability.

According to our goal, we set $\ket{\psi_{\rm tar}}=\ket{{\rm GHZ}}$.
Note that the target is composed of two eigenstates of the drift Hamiltonian with identical eigenvalues.
Therefore, once the GHZ state has been prepared, the control fields can be switched off, and the state  evolves trivially with the drift Hamiltonian.
Hence, the control problem fits well into the category of eigenstate preparation.

The coefficients $c_\mu$ from Eq.~\eqref{Eq:FundLoss} were obtained using the BOHB algorithm.
The same algorithm was also used to optimize other hyperparameters, i.e., the learning rate, the number of parallel trajectories $b$, and the number of weights in each layer of the NN for any size of the system $M$. 
Their values are summarized in Tab.~\ref{TabHyperPars} in~\ref{Ap:Table}.

\begin{figure}
	\centering
	\includegraphics[width=1\linewidth, angle=0]{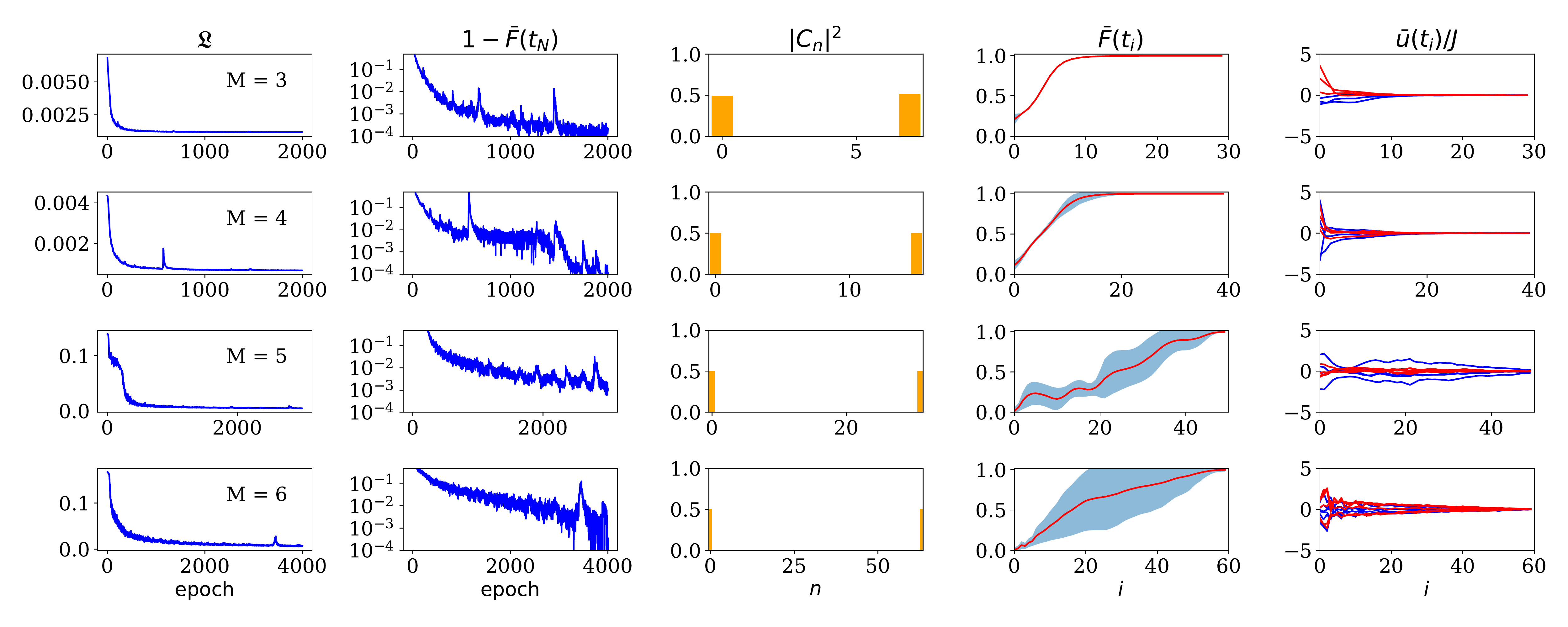}
	\caption{ GHZ state preparation for various lengths  $M$ of the spin chain.
	From top to bottom, the number of sites $M$ increases. 
		The first two columns show the progress of the loss function $\mathfrak{L}$ and the final state infidelity averaged over $b$ parallel trajectories $1-\bar{F}(t_N)$  as a function of the training epochs. 
		Columns three to five show results of the trained NN applied to a test set of 256 spin configurations.  
		The coefficients $C_n$ plotted in the third column refer to Eq.~\eqref{Eq:GenVecSpinChain}.
		The GHZ state corresponds to  equal occupations of the leftmost and the rightmost columns in the histogram, i.e., $\abs{C_0}^2=\abs{C_{D-1}}^2=0.5$.
		The evolution of the mean fidelity $\bar{F}$ of the test set during the preparation process for each step $i=1\ldots N$ is plotted in the fourth column. 
		The blue bands highlight one standard deviation.
		The last column depicts the applied $2M$ independent actions averaged over the test set for each step $i$. 
		 All hyperparameters can be found in Tab.~\ref{TabHyperPars} in~\ref{Ap:Table}.}
	\label{fig_GHZ}
\end{figure}

The training phase of the NN as well as examples of its performance on test data sets are shown in Fig.~\ref{fig_GHZ}.
As compared to a typical RL case, the loss functions $\mathfrak{L}$ show rapid initial progress and do not display any strong oscillations during the learning process.
The performance we require from our control scheme is to reach fidelities above $99.9 \%$ at latest at the end of the control interval.
Obviously, the number of the training epochs required to accomplish this grows with $M$ (see the second column of Fig.~\ref{fig_GHZ}). 
However, for all system sizes considered we manage to train the NN to reach the desired performance as demonstrated in the third and the fourth columns of Fig.~\ref{fig_GHZ}.
Indeed, even for the case with $M=6$ qubits, where the initial noise leads to higher fluctuations in the test data, the final average fidelity $\bar{F}(t_N)$ meets our precision requirements  while having effectively no dispersion at the same time.
The last column in Fig.~\ref{fig_GHZ} shows the averaged signals from the control fields.
The NN also learned to set all control fields to zero, once the GHZ state has been reached.  

\subsection{Eigenstate preparation in a quantum parametric oscillator}
\label{Subsec:EigPrepPar}
As the last example, we aim at the preparation of a specific eigenstate of the drift Hamiltonian from Eq.~\eqref{Eq:Parametron}.
The initial state is, again, considered as noisy, 
\begin{equation}
\ket{\psi(t_0)}=\ket{0}+ \sum_{n=0}^{ D-1}{\rm e}^{-\frac{n}{3}}\xi_n\ket{n}.
\label{Eq:iniState}
\end{equation}
In our numerical implementation we truncate the Hilbert space of photons to a finite dimension $D$ by  considering only the first $D$ Fock states  $\ket{n}$.
Random noise $\xi_n$  is uniformly sampled from an interval $[-\xi,\xi]$ where $\xi$ is a fixed parameter of the environment.
The exponential factor guarantees that the contribution of the highest-lying Fock states is reduced, i.e., $\ket{\psi(t_0)}$ is dominantly distributed among the low-lying Fock states.

Let $\ket{\alpha}$ be a coherent state  with complex $\alpha$~\cite{glauber1963}.
These states are often referred to as \lq the most classical ones\rq\ as they satisfy the minimal Heisenberg uncertainty relation.
We consider an example of a Schr\"{o}dinger cat state defined as $\ket{{\rm cat}_\alpha}=\frac{1}{\sqrt{2}} \left(\ket{\alpha} +\ket{-\alpha}\right)$~\cite{leibfried2005}.
The respective Wigner quasiprobability distribution  is formed by two  Gaussians in phase space located at the coordinates $(x=\sqrt{2} \ {\rm Re}\alpha, p=\sqrt{2} \ {\rm Im}\alpha)$ with a non-classical interference pattern between them.
Such states can serve as resources for quantum computing, since they can encode a qubit protected against phase-flip errors~\cite{cochrane1999,mirrahimi2014,grimm2019}.
In our setting of the model, a specific cat state with $\alpha=2$ happens to be the first excited eigenstate of the drift Hamiltonian.
We further refer to this state as the \textit{eigen-cat state}.
The goal of this section is to transfer the initial state $\ket{\psi(t_0)}$ to the target $\ket{\psi_{\rm tar}}=\ket{{\rm cat}_2}$.

\begin{figure}
	\centering
  \includegraphics[width=0.85\linewidth, angle=0]{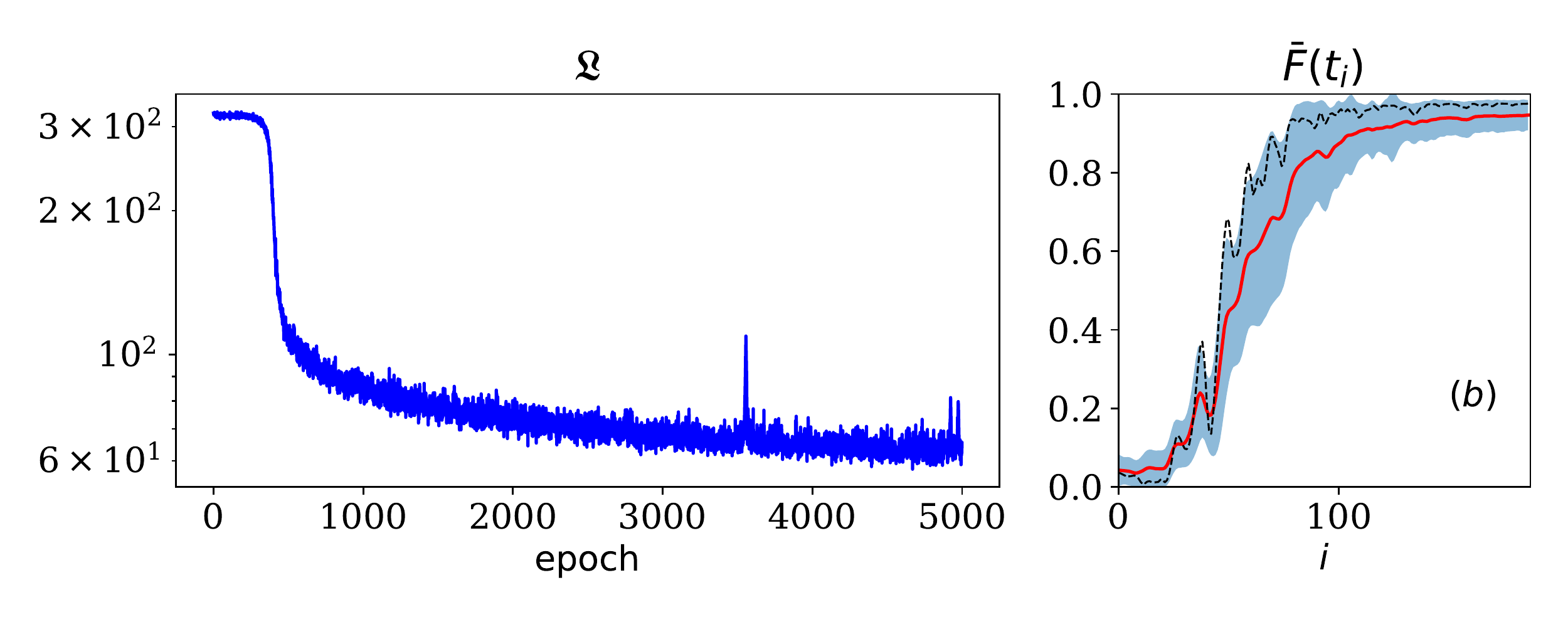} 
  \includegraphics[width=0.85\linewidth, angle=0]{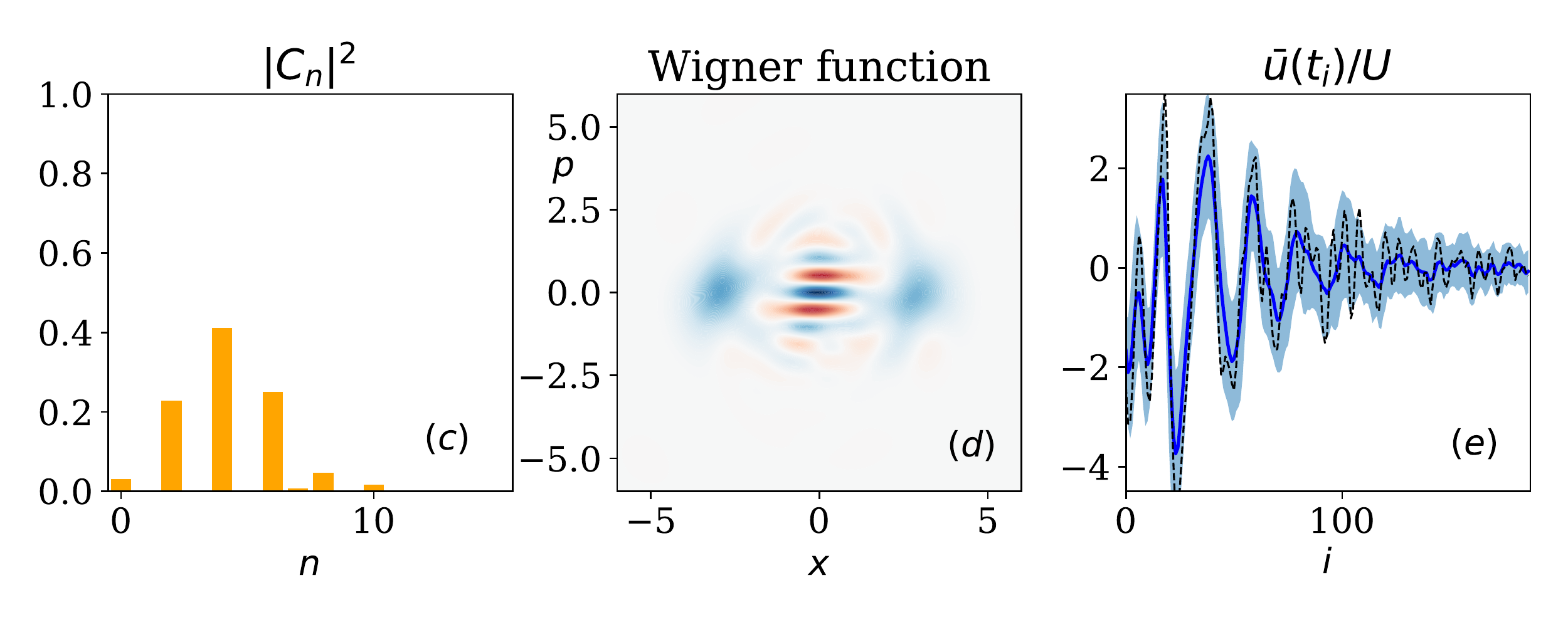}
	\caption{ Eigen-cat state preparation in the quantum parametric oscillator.
	The training of the NN and the performance on a test set with 64 initial states are depicted. The noise bandwidth is $2\xi=0.8$.
 Panel (a): Progress of the loss function $\mathfrak{L}$ during the training averaged over $b$ parallel trajectories.
 Panel (b): Evolution of the mean fidelity $\bar{F}$ of the test set, the blue region shows one standard error. 
 Panels (c) and (d): Fock distribution  and the Wigner function of a randomly chosen state from the test set at the final time step $t_N$.
 The coefficients $C_n$ refer to Eq.~\eqref{Eq:ParaState}.
Panel (e): Mean actions performed by the agent during the test run, the blue region shows one standard error.
The black dashed lines in panels (b) and (e) show a the special case with an unperturbed initial vacuum state $\ket{\psi(t_0)}=\ket{0}$.
All hyperparameters can be found in Tab.~\ref{TabHyperPars} in~\ref{Ap:Table}. }
	\label{fig_Cat}
\end{figure}

The loss function $\mathfrak{L}$, again, has the form prescribed by Eq.~\eqref{Eq:FundLoss}.
Together with other hyperparameters, the coefficients $c_\mu$ are presented in Tab.~\ref{TabHyperPars} in~\ref{Ap:Table}.
In this case, they were tuned empirically to achieve a desirable performance of the agent.\footnote{Hyperparameter optimization like BOHB could also be employed.
However, to create a probabilistic model a relatively large minimum budget is required such that the task is computationally expensive.}
The results are collected in Fig.~\ref{fig_Cat}.
As shown in panel~(a), after around 300 initial epochs, the agent starts learning quickly.
The learning process is stable without any significant oscillations.
The performance of the trained NN is shown in panels (b)--(e).
The mean fidelity of the test set reaches  $\bar{F} \approx 0.93$ with a small variance.
The state at the end of the control interval of a randomly chosen example from the test set is depicted  in panels~(c) and~(d).
The respective pulse applied is shown in panel~(e).
The pulse oscillates nearly symmetrically around zero which translates to oscillatory tilting of the classical potential to left and right, cf. Fig.~\ref{fig_spinchain}(c).
This qualitative behavior reflects the \lq symmetry\rq\ of the control task where the target state is equally distributed in both wells and the initial one (though slightly perturbed) at the top of the barrier.
As the fidelity approaches its final value, the amplitude of the control field decreases until it only oscillates noisily around zero.

Let us compare to a known protocol to prepare an eigen-cat state; it is based on an adiabatic attenuation of the amplitude $G$  of the parametric drive from Eq.~\eqref{Eq:Parametron}, see Ref.~\cite{pechal2019} for recent experimental result.
While this is also easily revealed by our agent, we assume here the amplitude $G$ to be a given constant over time and allow the agent only to control the amplitude of an additional single-photon drive.
For this problem, the exact pulse shape is not easily anticipated, but
our agent managed to find a non-trivial sequence with final fidelities comparable to Ref.~\cite{pechal2019}.

\section{Discussion and outlook}

For the systems tested, the differentiable programming method showed promising results and reliably found  successful protocols for eigenstate preparation.
The optimal strategies can be obtained after a few hundreds of training epochs.
The method is intrinsically robust towards uncertainties in the input data, thus provides more versatility compared to standard quantum control algorithms.
Also, the convergence is smooth and stable with respect to different initial seeds and small variations of the hyperparameters.

Despite the uncertainty in the initial states, we managed to reach fidelities larger than $99.9\%$ in the preparation of a GHZ state in a chain of $M$ qubits.
We exclusively used fully-connected linear layers, which limits the depth and thus also the representational power of the NN. 
Therefore, our current architecture is not well suited for high dimensional (more than $1000$) inputs.

In the case of the quantum parametric oscillator, the eigen-cat state preparation was accomplished with high fidelities despite the noisy input data.
In our current setting, the agent has  full information about the state at each time step.
To better represent experimental reality, some form of photon field measurement could be included.
The evolution of such an open system can be described by a stochastic Schr\"{o}dinger equation (SSE).
Optimal control with stochastic dynamics has drawn attention recently, see Refs.~\cite{abdelhafez19, krastanov19}.
Our method can be applied to open systems in a straightforward way by replacing the current ODE by a SSE.

In contrast to \textit{in situ} approaches \cite{ferrie2015},  we require a model of the physical system to set up the control tasks. 
However, recent advantages in parameter estimation from data bear the potential to bridge this gap~\cite{krastanov19, flurin2018}.
Furthermore, a student/teacher approach, similar to Ref. ~\cite{fosel2018}, is a conceivable solution.

\section{Conclusion}

In  summary,  we  have  introduced  a differentiable programming method for quantum control that leverages   information from the gradient obtained by differentiation through the dynamical equations of the system. 
The approach was successfully demonstrated on a single- and a many-body quantum system. 
The method is intrinsically robust towards uncertainty in the input states.
Moreover, its application to open quantum systems is straightforward.
Due to these attributes, we hope it will become a useful part of physicists’ toolbox to control complex quantum systems.


\section*{Acknowledgment}
This work was financially supported by the Swiss National Science Foundation (SNSF) and the NCCR Quantum Science and Technology. Calculations were performed at sciCORE (scicore.unibas.ch) scientific computing core facility at University of Basel.

\section*{References}
\bibliography{DPQC_bibfile}

\appendix

\section{Hyperparameters of the models}
\label{Ap:Table}
The hyperparameters used in the control tasks are summarized in Tab.~\ref{TabHyperPars}.

\begin{table*}
	\centering\ra{1.3}
	\resizebox{\textwidth}{!}{%
		\begin{tabular}{@{}rrcrrrrcrr@{}}\toprule
			& \multicolumn{1}{c}{parametric osc.} & \phantom{abc}& \multicolumn{4}{c}{spin chain} & \phantom{abc} & \multicolumn{2}{c}{qubit}\\
			\cmidrule{2-2} \cmidrule{4-7} \cmidrule{9-10} 
			& eigen-cat   && $M=3$ & $M=4$ & $M=5$ &$M=6$ && multiple losses & single loss\\\midrule
			\multicolumn{1}{l}{Hilbert space}\\
			$D$ & 16 && 8 & 16 & 32 & 64 && 2 & 2 \\
			\multicolumn{1}{l}{ODE solver}\\
			$N$ & 187 && 30 & 40 & 50 & 60 && 150 & 150 \\
			$N_{\rm sub}$ &  200 && 20 & 20 & 20 & 20 && 20 & 20 \\
			$\der t$ & $10^{-4}U$  && $0.001J$ & $0.001J$ & $0.001J$ & $0.001J$ && $0.01 \omega$ & $0.01 \omega$ \\
			\multicolumn{1}{l}{loss function}\\
			$\gamma$ & 0.999 && 1.0 & 1.0 & 1.0 & 1.0 && 1.0 & 1.0 \\
			$c_F$ & 0.8 && $1.8\cdot10^{-4}$ & $1.1\cdot10^{-4}$ &$2.0\cdot10^{-4}$  & $3.0\cdot10^{-4}$ && 0.57 & $3.1\cdot10^{-4}$ \\
			$c_{FN}$ & 200  &&$8.1\cdot10^{-6}$  & $3.6\cdot10^{-7}$ &$0.13$  & $0.15$ && $2.6\cdot10^{-3}$ & 0 \\
			$c_{\rm amp}$& 0.01 && 0 & 0 & 0 & 0 && 0 & 0\\
			$c'_{\rm amp}$& 0 &&  $4.2\cdot10^{-5}$ & $4.1\cdot10^{-6}$ & $1.7\cdot10^{-6}$  & $1.7\cdot10^{-6}$ && $3.9\cdot10^{-6}$ & 0\\			
			\multicolumn{1}{l}{Adam} \\
			learning rate & $4\cdot10^{-5}$ && $8.4\cdot10^{-4}$ & $7.0\cdot10^{-4}$ &  $6.0\cdot10^{-4}$ &  $6.0\cdot10^{-4}$ && $3.3\cdot10^{-3}$ & $4.9\cdot10^{-4}$ \\
			$b$ & 64 && 256 & 512 & 256 & 256 && 256 & 256 \\
			epochs & 3000  && 2000 & 2000 & 3000 & 4000 && 400 & 400 \\ 
			\multicolumn{1}{l}{NN  }\\
			LLS 1 & (32, 512) && (16, 512) & (32, 256) &  (64, 128) & (128, 256) && (4, 256) & (4, 256) \\
			LLS 2 & (512, 256) && (512, 32) & (256, 256) &  (128, 128) & (256, 256) &&(256, 256) & (256, 256) \\
			LLS 3 & (256, 256) && \phantom{(4, 256)} &  \phantom{(128, 128)} & (128, 128) & (256, 256) && (256, 128) & (256, 128) \\
			LLS 4 & (256, 64) && \phantom{(256, 64)} &  \phantom{(256, 64)} & \phantom{(256, 64)} & \phantom{(256, 64)} && \phantom{(256, 64)} & \phantom{(256, 64)} \\
			LLA 1 & (2, 128) && (6, 16) & (8, 64) & (10, 32) & (12, 64) &&  (1, 128) & (1, 128) \\
			LLA 2 & (128, 64) && (16, 32) & (64, 256) &  (32, 32) & (64, 64) &&(128, 128) & (128, 128) \\
			LLA 3 & \phantom{(128, 64)} &&  \phantom{(4, 256)} & \phantom{(4, 256)} &  (32, 128) & (64, 256) && \phantom{(128, 128)} & \phantom{(128, 128)}\\
			LLC 1 & (64, 64)  && (32, 32) & (256, 256) & (128, 128) & (256, 256) && (128, 64) & (128, 64) \\
			LLC 2 & (64, 32) && (32, 6) & (256, 8) & (128, 128) & (256, 256) && (64, 32) & (64, 32) \\
			LLC 3 & (32, 2) &&  \phantom{(4, 256)} & \phantom{(128, 10)} & (128, 10) & (256, 12) && (32, 1) & (32, 1) \\		
			\bottomrule
	\end{tabular}}
	\caption{Hyperparameters employed to obtain the results in the main text. \lq LLS~1\rq\ stands for first linear layer state-aware network, etc.}
	\label{TabHyperPars}
\end{table*}

\section{Hyperparameter optimization}
\label{Ap:BOHB}

\begin{figure}
     \centering
     \includegraphics[width=0.495\linewidth, angle=0]{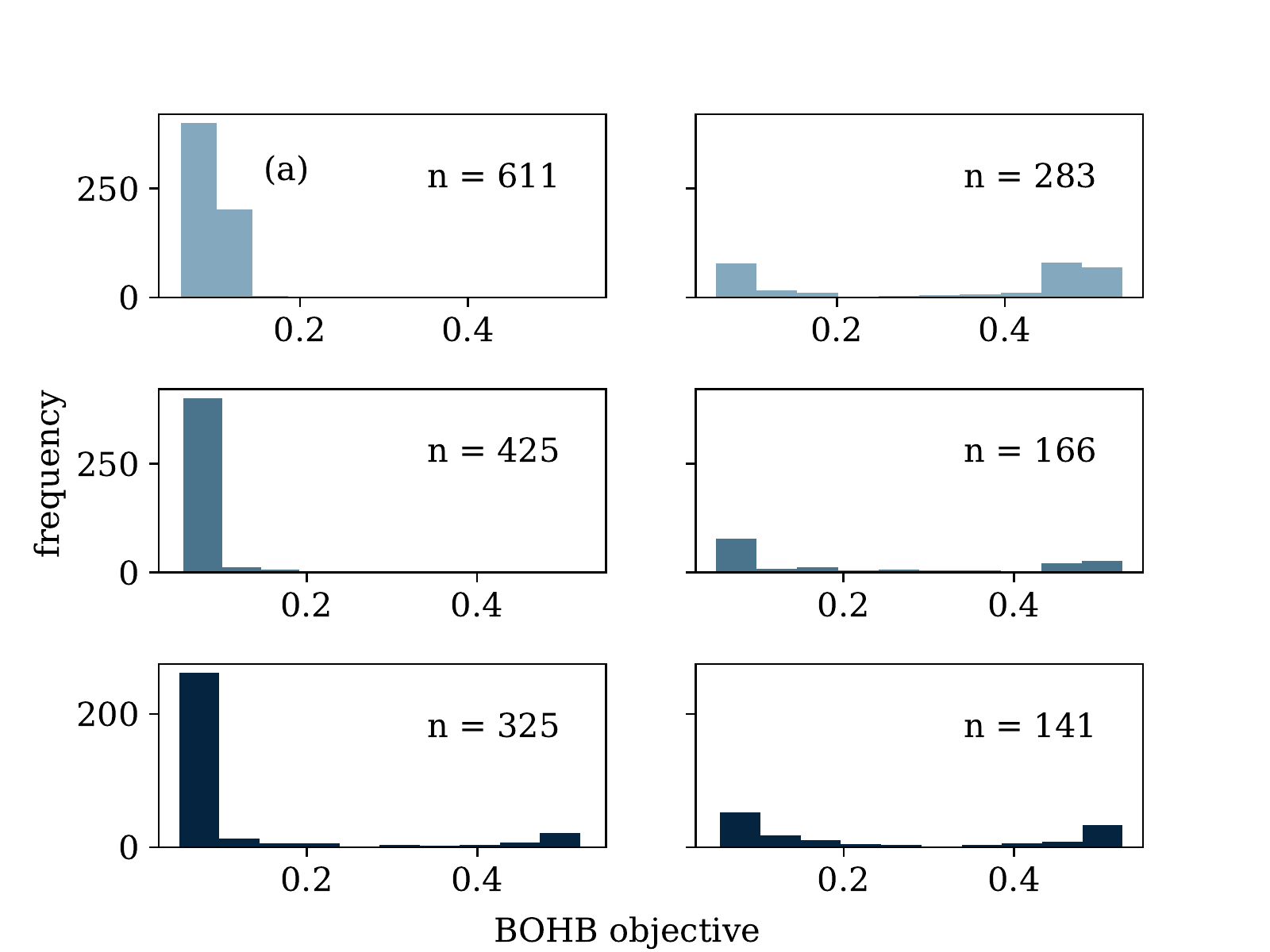}
     \includegraphics[width=0.495\linewidth, angle=0]{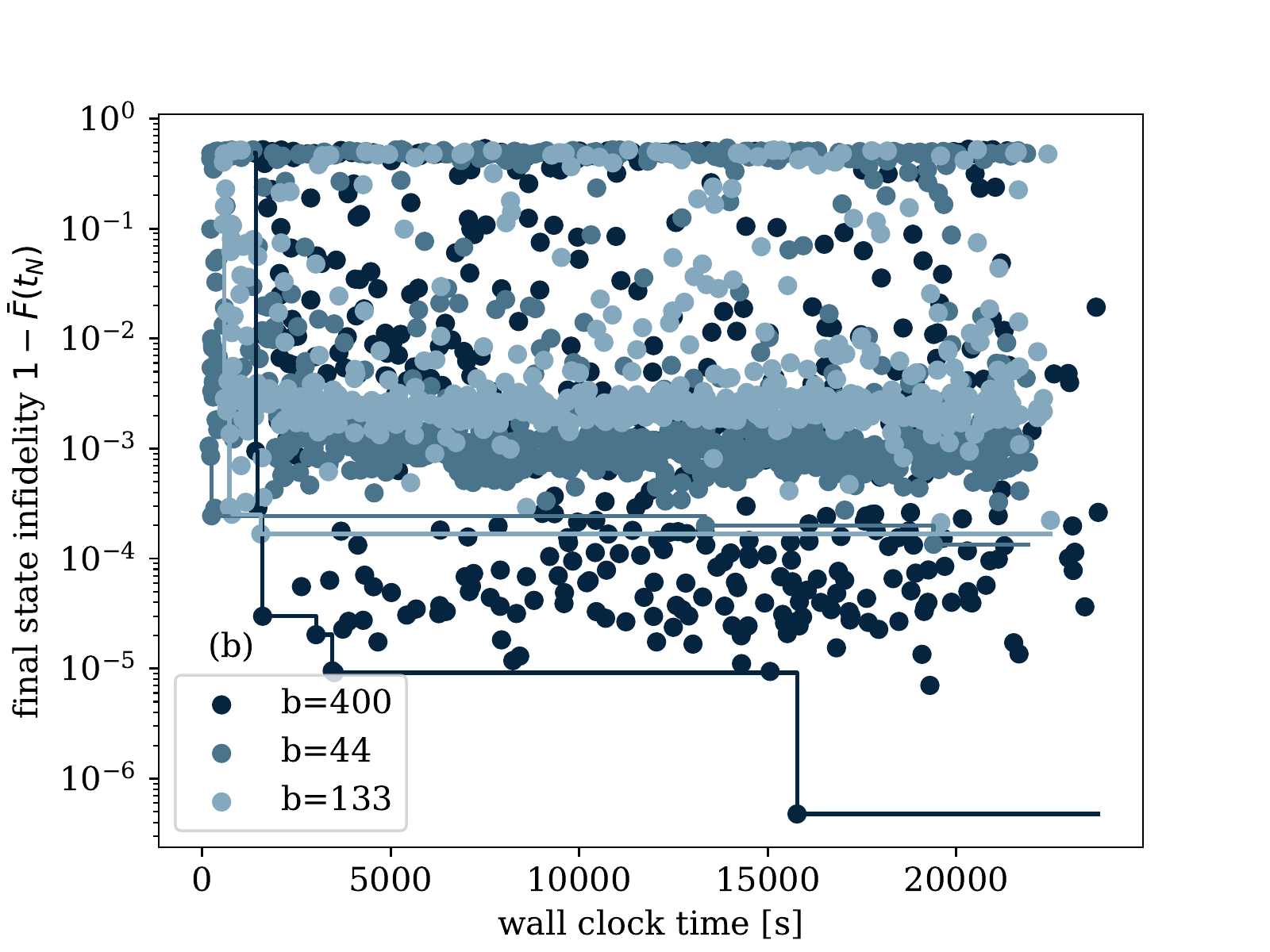}
     \caption{
         BOHB optimization of the hyperparameters for the manipulation
of a single qubit~\eqref{eq_qubit} with figure of merit chosen as
$\mathfrak{L} = c_F \mathfrak{L}_F$ (see Eq.~\eqref{Eq:FundLoss}).
         Panel (a): Frequency distribution as a function of the BOHB objective
for model-based picks (left) versus random-configurations (right) for
three budgets (rows).
         The number $n$ in each cell indicates the total number of
samples that fall into the category.
         Panel (b): Final state infidelity, $1-\bar{F}(t_N)$, as a
function of the duration of the simulation (total wall time).
         The budgets -- and data -- are identical to (a), as highlighted
by the color code.   }
     \label{fig_qubit_bohb}
\end{figure}

In this appendix, we illustrate the results obtained by the hyperparameter optimization with the Bayesian Optimization and Hyperband method (BOHB)~\cite{falkner2018} in the case of the qubit control task with $\mathfrak{L}=c_F \mathfrak{L}_F$ (see Eq.~\eqref{Eq:FundLoss}).
BOHB is based on the well-known Hyperband  approach~\cite{li2017} to determine how many hyperparameter configurations are evaluated at a certain budget, c.f. pseudocode in Ref.~\cite{falkner2018}.
In our case, the budget is identical to the number of epochs that are used to train the predictive model.
In contrast to Hyperband, BOHB replaces the random selection of configurations at the beginning of each iteration by a model-based search.
Figure~\ref{fig_qubit_bohb}(a) visualizes the frequency distribution as a function of the BOHB objective, $\mathfrak{L}_F$, for model-based and random configurations.
As desired, nearly all model-based picks achieve a very small value for $\mathfrak{L}_F$. 
For both distributions, an increase of the budget leads to smaller values of $\mathfrak{L}_F$ on average, as expected.
The differences between the three budgets can also be seen in the evolution of the final state infidelity, $1-\bar{F}(t_N)$, as a function of the wall clock time in Fig.~\ref{fig_qubit_bohb}(b).
Note that, the infidelity is plotted in log-scale and that many configurations reach a very high fidelity.

\section{Differentiable programming versus REINFORCE}
\label{Ap:CompareRLvsDP}
Here we compare the performance of our differentiable programming (DP) method with a vanilla policy gradient algorithm (REINFORCE)~\cite{sutton1998,williams1992,  sutton2000} in the case of the qubit control problem from Section~\ref{Sec:Qubit}.
In a nutshell, our REINFORCE implementation is based on three substeps:
Firstly, trajectories $\tau$ are sampled from the current Gaussian policy
\begin{equation}
\log \pi_\theta(u_x(t_i)|\left\{\ket{\psi(t_i)}, u_x(t_{i-1})\right\}) =
- \frac{1}{2}\left( \frac{(u_x(t_i) - \mu)^2}{\sigma^2} +\log(2 \pi
\sigma^2) \right),
\end{equation}
where the variance $\sigma^2=0.04$ is fixed and the mean $\mu = \mu_\theta(\left\{\ket{\psi(t_i)}, u_x(t_{i-1})\right\})$ is determined by the predictive model, which is structured identically to the one in the DP case, see Tab.~\ref{TabHyperPars} in~\ref{Ap:Table}.
Secondly, based on the rewards-to-go
\begin{equation}
R_n = \sum_{n' = n}^N \left[ c_F |\langle \psi(t_{n'})|
\psi_{\rm tar}\rangle|^2  - c'_{\rm amp} |u_x(t_n)|^2 + \delta_{n',N}
c_{FN}|\langle \psi(t_{n'})| \psi_{\rm tar}\rangle|^2\right],
\end{equation}
where the (properly adjusted) elementary loss functions from Section~\ref{Sec:Loss} are employed, we approximate the gradient of the expected total reward over all trajectories $J_{\theta}$ as
\begin{equation}
\nabla_\theta J_\theta = \underset{\tau \sim \pi_\theta}{\rm{E}}
\left[\sum_{n=0}^N \nabla_\theta \log
\pi_\theta(u_x(t_n)|\left\{\ket{\psi(t_n)}, u_x(t_{n-1})\right\})R_n
\right].
\end{equation}
Note that while the loss functions for the DP are minimized, the rewards in RL are maximized. Note further that the rewards-to-go are normalized before they enter the approximation of the gradient~\cite{karpathy16}.
Thirdly, we use the ADAM optimizer with a learning rate of $0.00025$ to update the network parameters in a series of epochs, each consisting of $1024$ trajectories.

\begin{figure}[h!]
     \centering
     \includegraphics[width=1\linewidth, angle=0]{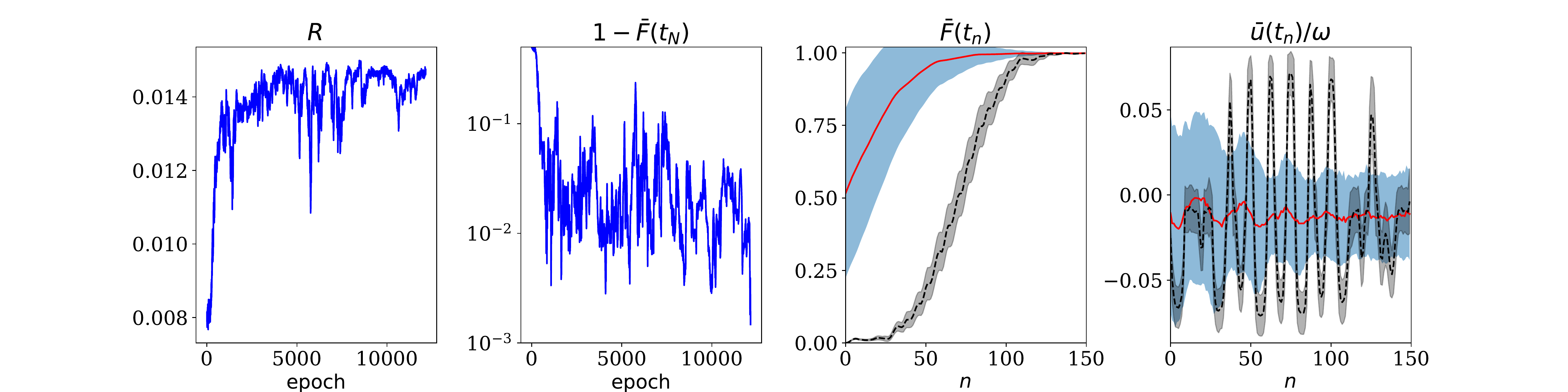}
     \caption{
         Preparation of the eigenstate  $\ket{\uparrow}_z$ of the drift term of the Hamiltonian~\eqref{eq_qubit} for the optimized hyperparameters containing the coefficients $c_F$, $c_{FN}$, $c'_{\rm amp}$ based on a REINFORCE implementation.
         In the first column, the evolution of the mean reward $R$ during the training phase is shown. 
         The second column shows the mean value of the final state infidelity as a function of the epochs.
         The third and fourth columns show the performance of the model when it is applied to a test set of  size 512.
         The red curve/ blue shaded region indicate the mean/standard deviation of the fidelity and the actions, respectively.
         The black dashed line highlights the special case with initial state given as $\ket{\psi(t_0)}=\ket{\downarrow}_z$.
     }
     \label{fig_qubit_RL}
\end{figure}

Figure~\ref{fig_qubit_RL} is analogous to  Fig.~\ref{fig_qubit} with RL replacing DP.
In general, we obtain a noisier evolution of the total expected reward during training as compared to the (relatively) smooth appearance of the training loss in the  main text.
Furthermore, the outcome is more sensitive to small changes of hyperparameters including the employed random seed.
We expect that more sophisticated approaches, such as trust-region policy optimization~\cite{schulman2015} or proximal policy optimization~\cite{schulman2017}, in combination with generalized advantage estimation~\cite{schulman2015GAE} could improve the performance.
As expected, more epochs were needed to train the neural network.
Also, the agent did not learn to push the control fields towards zero once the target state has been prepared.

\begin{figure}[h!]
     \centering
     \includegraphics[width=1.0\linewidth, angle=0]{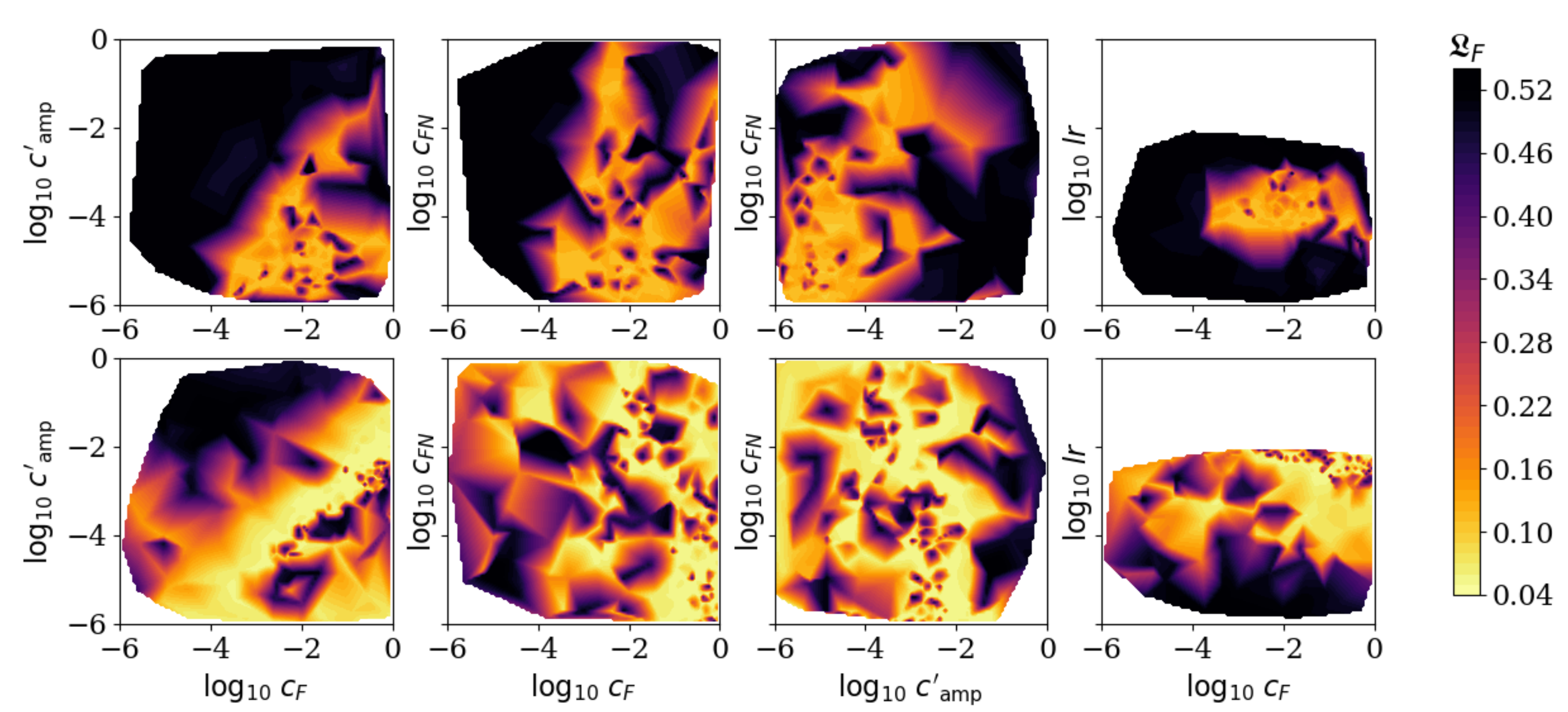}
     \caption{
         Optimal regions in the hyperparameter space consisting of $c_F$, $c'_{\rm amp}$, $c_{FN}$ and learning rate $lr$ of ADAM as established by BOHB for the DP approach (bottom row) and the RL approach (top row).
         The figures show projections from the four-dimensional space onto the respective two-dimensional subspaces.
         The color map encodes the main objective of BOHB to be minimized, i.e., the average infidelity over the entire time interval $\mathfrak{L}_F$.
         The brighter areas correspond to a better performance.
         The continuous map results from an interpolation between the discrete set of sampled configurations probed by the BOHB algorithm.
     }
     \label{fig_OptHyper}
\end{figure}

A comparison between  the mean infidelity $\mathfrak{L}_F$ over all time steps as a function of the hyperparameters $c_F$, $c_{FN}$, $c'_{\rm amp}$ and the learning rate $lr$ for both approaches is depicted in Fig.~\ref{fig_OptHyper}.
What is plotted are the two-dimensional projections of $\mathfrak{L}_F$ onto the subspaces of the respective pairs of hyperparameters.
The fact that the landscapes for the DP method (lower panel) reach lower infidelities as compared to the REINFORCE implementation (top panel)  shows that the DP method generally performs better.
Also the bright regions, indicating a successful performance, are more extended in our approach which implies that it is more stable with respect to changes in the hyperparameters.
Remarkably, DP is trained only for 400 epochs while 10000 epochs are used in case of the REINFORCE algorithm.

\end{document}